\documentclass[runningheads]{llncs}
\usepackage{graphicx}
\usepackage[utf8]{inputenc}
\usepackage{textcomp}
\usepackage{graphicx,verbatim}
\usepackage{hyperref}
\usepackage{cite}
\usepackage{amsmath}
\usepackage{amssymb}
\usepackage{authblk}
\usepackage{bbding}
\usepackage{graphicx}
\usepackage{booktabs}
\usepackage{graphicx}
\usepackage[normalem]{ulem}
\usepackage{multirow}
\usepackage{amsmath} 
\usepackage{bm}      
\usepackage{multirow}
\usepackage[table,xcdraw]{xcolor}
\usepackage{colortbl}
\usepackage[normalem]{ulem}
\useunder{\uline}{\ul}{}
\usepackage[ruled,vlined]{algorithm2e}
\usepackage{algpseudocode} 
\begin{document}
\title{Beyond Pixel Agreement: Large Language Models as Clinical Guardrails for Reliable Medical Image Segmentation}
\titlerunning{Hierarchical Clinical Reasoner}

\author{Jiaxi Sheng\inst{1,2}, Leyi Yu\inst{3}, Haoyue Li\inst{1,5}, Yifan Gao\inst{1,2,4}\textsuperscript{\Envelope}, Xin Gao\inst{2,4}\textsuperscript{\Envelope} }
\authorrunning{J. Sheng et al.}
\institute{
	School of Biomedical Engineering (Suzhou), Division of Life Science and Medicine, University of Science and Technology of China, Hefei, China \and
	Suzhou Institute of Biomedical Engineering and Technology, Chinese Academy of Sciences, Suzhou, China \and
	Department of Electrical, Computer, and Systems Engineering, Case Western Reserve University, Cleveland, OH, USA \and
	Shanghai Innovation Institute, Shanghai, China \and
	College of Medicine and Biological Information Engineering, Northeastern University, Shenyang, China \\
}

\maketitle              
\begin{abstract}
Evaluating AI-generated medical image segmentations for clinical acceptability poses a significant challenge, as traditional pixel-agreement metrics often fail to capture true diagnostic utility. This paper introduces Hierarchical Clinical Reasoner (HCR), a novel framework that leverages Large Language Models (LLMs) as clinical guardrails for reliable, zero-shot quality assessment. HCR employs a structured, multi-stage prompting strategy that guides LLMs through a detailed reasoning process, encompassing knowledge recall, visual feature analysis, anatomical inference, and clinical synthesis, to evaluate segmentations. We evaluated HCR on a diverse dataset across six medical imaging tasks. Our results show that HCR, utilizing models like Gemini 2.5 Flash, achieved a classification accuracy of 78.12\%, performing comparably to, and in instances exceeding, dedicated vision models such as ResNet50 (72.92\% accuracy) that were specifically trained for this task. The HCR framework not only provides accurate quality classifications but also generates interpretable, step-by-step reasoning for its assessments. This work demonstrates the potential of LLMs, when appropriately guided, to serve as sophisticated evaluators, offering a pathway towards more trustworthy and clinically-aligned quality control for AI in medical imaging.

\keywords{Large Language Models  \and Medical Image Segmentation \and Automatic Quality Control \and Hierarchical Clinical Reasoner.}
\end{abstract}
\section{Introduction}
Medical image segmentation, which partitions images into meaningful regions, provides essential quantitative information for disease diagnosis, treatment planning, and monitoring \cite{gao2025wega,xu2025multi,xu2025tooth}. The precise delineation of anatomical structures and pathologies, such as tumors or organs-at-risk, directly informs clinical decision-making. Advances in artificial intelligence (AI), particularly deep learning \cite{dai2021transmed,xu2024facial}, have significantly improved the automation, accuracy, and efficiency of these segmentation tasks across various imaging modalities and applications \cite{gao2023anatomy,gao2024desam,gao2024mba,yuan2025abs}. AI-driven systems can alleviate the laborious manual contouring efforts of physicians, offering potential for reduced workload and enhanced consistency. However, deploying these AI systems in routine clinical practice presents challenges, primarily concerning their reliability. While often accurate, models can produce erroneous segmentations, and undetected errors may lead to severe consequences, including misdiagnosis or inappropriate treatment.

The conventional quality assurance method, involving manual expert verification of each AI-generated segmentation, is time-consuming and subjective, undermining the efficiency benefits of AI automation \cite{azad2024medical}. This bottleneck is particularly acute in high-throughput clinical environments where large volumes of images are processed daily or where rapid results are needed. The variability in model performance can be attributed to factors such as diverse image acquisition protocols, complex anatomical structures, ambiguous lesion boundaries, and domain shifts. This situation creates an "efficiency paradox": AI aims to boost efficiency, but the validation needed due to potential errors can counteract these gains if performed manually. Therefore, automated quality control (QC) is a requirement to realize the full potential of segmentation in demanding clinical settings, encompassing not just high volume but also scenarios requiring swift processing for urgent decisions or handling complex cases. \cite{van2022qualitative,lin2024no}

Current automated QC approaches often involve training secondary AI models to predict quantitative metrics like the dice similarity coefficient \cite{fournel2021medical,zhang2023sqa,uslu2024robust,zaman2023segmentation}. These methods typically use the original image and the AI-generated mask as input, sometimes augmented with uncertainty or error maps. While these techniques show promise in providing continuous quality scores or flagging failed cases via thresholding, their correlation with true clinical acceptability can be imperfect. Metrics like dice similarity coefficient, while widely used, may not always capture clinically relevant errors, such as small but significant omissions or imprecise boundary adherence in critical regions. Other QC strategies leverage uncertainty quantification or direct error detection \cite{lambert2024trustworthy,zhao2022efficient}, but translating these outputs into direct, actionable clinical usability categories remains an area of active development. There is an evident need for evaluation mechanisms that extend beyond pixel-level agreement to better align with a clinician's interpretation of segmentation reliability for patient care.

\begin{figure*}[htbp]
	\centering
	\includegraphics[width=0.9\textwidth]{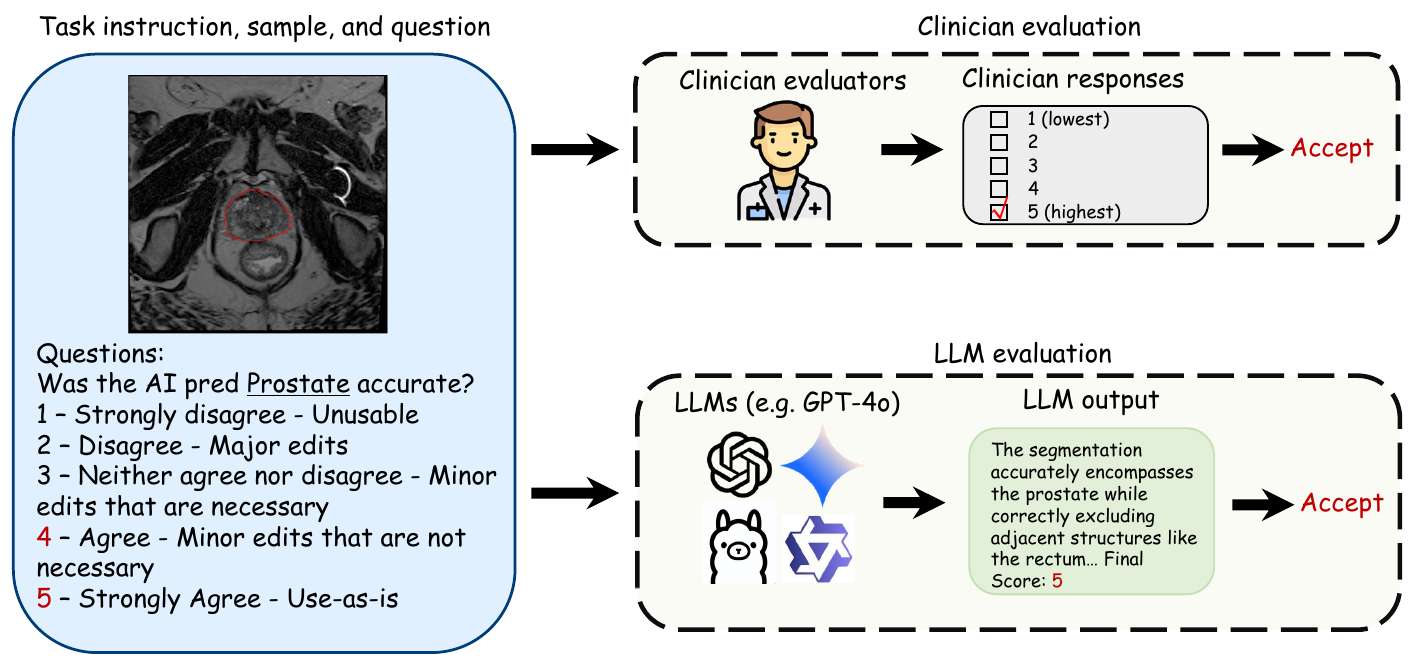}
	\caption{Conceptual overview of the medical image segmentation quality assessment task addressed in this work. The illustration outlines (a) the core components presented for evaluation, including task instructions, an image sample, and a guiding question; (b) the traditional clinician-centric evaluation process; and (c) our proposed LLM-driven evaluation process facilitated by the Hierarchical Clinical Reasoner (HCR).}
	\label{fig:task_overview}
\end{figure*}

In this paper, we introduce Hierarchical Clinical Reasoner (HCR), a novel approach that employs large language models (LLMs) as clinical guardrails for the reliable evaluation of medical image segmentation. Figure~\ref{fig:task_overview} conceptually outlines this task, contrasting the evaluation pathway undertaken by human clinicians with the structured reasoning process we implement within our LLM-based HCR. We posit that LLMs, guided by a sophisticated, multi-stage prompting strategy, can perform assessments with high fidelity to clinical judgment, more so than traditional pixel-based metrics or simpler automated QC methods. Our HCR framework directs the LLM to engage in a step-by-step reasoning process, commencing with recalling modality-specific anatomical knowledge, progressing to an analysis of low-level visual features of the segmentation, then to anatomical inference, and culminating in a high-level clinical synthesis and scoring. The entire evaluation is outputted in a structured format, detailing the reasoning at each stage. On a diverse benchmark across six distinct medical imaging tasks (covering CT, MR, and PET-CT modalities), our proposed method, utilizing a model like Google's Gemini 2.5 Flash, achieved a quality classification accuracy of 78.1\%. This zero-shot performance, driven by structured prompting alone, is competitive with established vision architectures such as ResNet50 (72.9\% accuracy) and EfficientNet-B0 (71.9\% accuracy) that were specifically trained for this classification task. This work underscores the potential of LLMs to serve as sophisticated, interpretable, and clinically-aligned evaluators, paving the way for more trustworthy AI in medical image analysis.

\section{Related Work}

\subsubsection{Automated Quality Control for Medical Image Segmentation}
The automated assessment of medical image segmentation quality is an active area of research, driven by the need to ensure the reliability of AI-driven tools in clinical practice \cite{van2022qualitative,chen2024quality}. Current automated QC methodologies frequently involve training secondary models to predict quantitative performance metrics, such as the dice similarity coefficient, or utilize uncertainty quantification as an indicator of segmentation trustworthiness \cite{fournel2021medical,zhang2023sqa,uslu2024robust,zaman2023segmentation}. These approaches, alongside direct error detection techniques  and the use of thresholds on predicted metrics to categorize usability, aim to streamline the review process and lessen the burden on expert clinicians in high-volume settings. However, a persistent challenge is that established metrics based on pixel overlap do not always fully correspond with clinical acceptability, as they may not capture all error types pertinent to downstream clinical decision-making.

\subsubsection{Large Language Models}
Large Language Models (LLMs) have demonstrated transformative progress in natural language understanding, generation, and complex reasoning \cite{chang2024survey,thirunavukarasu2023large,liu2025application}. The recent development of multimodal LLMs has extended these capabilities to encompass visual data, allowing for joint processing and interpretation of images and text \cite{wu2024gpt,liang2025multimodal,hu2024bliva}. A significant line of inquiry in current LLM research focuses on eliciting advanced performance through sophisticated prompt engineering strategies, often targeting robust zero-shot generalization without requiring model fine-tuning for specific downstream tasks \cite{wang2023large,wang2024prompt}. While this evolution positions LLMs for tasks requiring expert-level judgment, the challenge of systematically applying them to generate detailed, clinically-aligned quality assessments of medical image segmentations has remained largely unaddressed. To our best knowledge, it is the first framework that leverages LLMs for automated quality control in medical image segmentation.

\section{Method}

\begin{figure}[htbp]
	\centering
	\includegraphics[width=\textwidth]{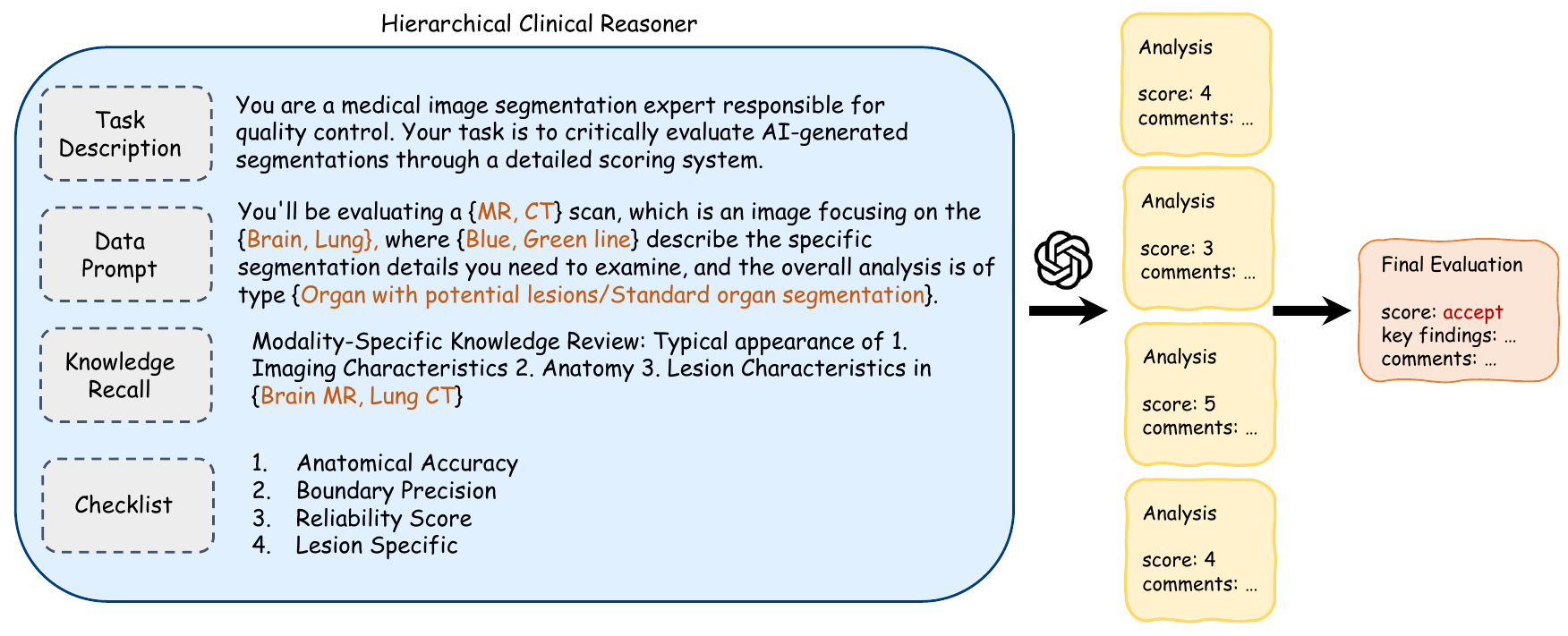}
	\caption{An overview of the Hierarchical Clinical Reasoner (HCR) framework, illustrating the multi-stage process from input (medical image with AI segmentation) to the structured clinical quality assessment generated by the LLM.}
	\label{fig:hcr_workflow}
\end{figure}

\subsection{Method Overview}
The goal of our Hierarchical Clinical Reasoner (HCR) is to provide a clinically-aligned quality assessment of AI-generated medical image segmentations using LLMs. Our method primarily consists of two components: the preparation of diverse medical image segmentation datasets for evaluation (Sec 3.2), and the HCR itself (Sec 3.3), where an LLM executes a multi-stage clinical reasoning process. This reasoning is guided by a structured text prompt and applied to the prepared visual data, enabling the LLM to output a comprehensive, interpretable evaluation culminating in a clinical usability recommendation.

\subsection{Dataset Preparation}
To robustly evaluate our HCR, the meticulous preparation of a suitable dataset comprising medical image segmentations with varying quality levels is essential. Merely utilizing an uncurated stream of AI-generated segmentations is insufficient for a robust assessment of an LLM's ability to perform detailed clinical quality evaluation. Such raw data often lacks the well-defined spectrum of quality and clear delineations of clinical acceptability required to rigorously probe these advanced evaluative capabilities. Our evaluation dataset originates from  \cite{murugesan2024ai}, a public repository providing AI-generated segmentations for numerous cancer types and imaging modalities. From these collections, we selected representative 2D image slices. Each selected 2D case presents the original medical image slice (e.g., Brain Lesion on MR scans or Lung Nodule on CT images) with the corresponding AI-generated segmentation overlaid. Subsequently, these cases were independently reviewed by clinical experts who assigned a ground truth quality label ("accept" or "reject") based on perceived clinical usability, yielding a curated collection of 479 image-label pairs for our experiments. Table~\ref{tab:dataset_overview} summarizes the composition of this dataset, detailing the distribution of samples across various imaging modalities, primary segmentation targets, and their allocation into training and testing sets.

\begin{table}[htbp]
	\centering
	\caption{Overview of the curated dataset for evaluating HCR, detailing the distribution of image samples across different anatomical site/modality groups, and primary segmentation targets.}
	\label{tab:dataset_overview}
	\begin{tabular}{cccccc}
		\toprule
		Dataset & Modality & Primary Target  & Total Cases \\
		\midrule
		brain-mr & MR & Brain Lesion & 48 \\
		breast-mr & MR & Breast Lesion & 90 \\
		liver-ct & CT & Liver & 53 \\
		lung-ct & CT & Lung Nodule & 172 \\
		lung-fdg-pet-ct & PET-CT & Lung Tumor & 35 \\
		prostate-mr & MR & Prostate & 81 \\
		\midrule
		\textbf{Total} & & & \textbf{479} \\
		\bottomrule
	\end{tabular}
\end{table}

\subsection{Hierarchical Clinical Reasoner}
Our proposed framework leverages LLMs to conduct a detailed and clinically-aligned quality assessment of AI-generated medical image segmentations. The overall HCR process, depicted in its entirety in Figure~\ref{fig:hcr_workflow}, is fundamentally orchestrated by a sophisticated, structured text prompt. This prompt acts as the primary interface, meticulously guiding the LLM through a multi-stage clinical reasoning pathway when it is presented with a 2D medical image that displays an AI-generated segmentation (prepared as detailed in Section 3.2).

The design of the structured text prompt is central to the HCR's efficacy. It is engineered to communicate several key pieces of information to the LLM: the specific clinical context, such as the target organ and imaging modality; a clear definition of its expert evaluator role; the precise multi-stage methodology it must follow; and the required format for its structured output. We found it important to explicitly instruct the LLM to ground its assessment in recalled anatomical and modality-specific knowledge before proceeding to visual feature analysis. Furthermore, the prompt is designed not merely to elicit a final score, but to compel the LLM to articulate its reasoning and observations at each stage of the evaluation, thereby ensuring transparency and interpretability.

The HCR's multi-stage reasoning pathway begins with a knowledge activation phase. In this initial step, the prompt directs the LLM to articulate its understanding of the typical appearance of the specified target organ and imaging modality, including expected tissue densities or signal intensities, common textures, and surrounding anatomical structures. This establishes a relevant baseline and contextual understanding for subsequent analysis. Following this, the LLM transitions to the low-level visual feature analysis stage. Here, it is guided to examine perceptually straightforward characteristics of the provided segmentation contour, such as its continuity as a complete loop, the presence of clear pixel intensity transitions at its interface with surrounding tissue indicative of visual edges, and the homogeneity or heterogeneity of the segmented region's internal texture.

These initial, more objective observations then inform the mid-level anatomical inference stage. At this juncture, the LLM assesses the segmentation's anatomical plausibility by comparing the visual features of the contour and the region it delineates against the recalled anatomical knowledge from the first stage. It evaluates whether the segmentation respects known anatomical boundaries and if its shape and location are consistent with the target organ or pathology. The LLM also evaluates the extent of target coverage, specifically identifying potential under-segmentation, meaning areas of the target missed by the contour, or spillage, referring to the erroneous inclusion of adjacent healthy tissues or different structures. This stage bridges the gap between raw visual data and clinical interpretation. The pathway concludes with the high-level clinical synthesis stage, where the LLM integrates all prior findings—knowledge recall, visual features, and anatomical inferences—into a coherent overall assessment of the segmentation's quality. This involves generating a concise clinical summary that highlights key strengths and weaknesses, assigning a numerical score from a predefined 1-5 scale reflecting clinical usability, and providing a descriptive category corresponding to this score. The resulting structured evaluation provides a detailed audit trail of the LLM's reasoning, offering clinicians not just a quality judgment but also an understanding of its basis.

\section{Results}

\subsection{Experimental Setup}
Our experiments were conducted on the curated dataset of 479 medical image segmentations, sourced as described in Section 3.2; the AI-generated segmentations under evaluation in our study were produced by the nnU-Net \cite{isensee2021nnu} framework. This dataset was partitioned into an 80\% training set (383 samples) for developing baseline models and a 20\% test set (96 samples) for evaluating all approaches, distributed across six distinct organ/modality groups (Table~\ref{tab:dataset_overview}). The primary task was to classify the quality of these generated segmentations into two clinical usability categories: "accept" or "reject". For this classification, segmentations originally rated by clinical experts as 4 or 5 on a 5-point usability scale were labeled as "accept", while those rated 1 to 3 were labeled as "reject". For comparative baselines, we trained three established vision architectures—EfficientNet-B0 \cite{tan2019efficientnet}, ResNet50 \cite{he2016deep}, and a Vision Transformer (ViT-Base) \cite{dosovitskiyimage}—on our training set. These models were tasked with directly classifying the quality from the input images, which displayed the original medical scan with the nnU-Net segmentation overlaid. Our proposed Hierarchical Clinical Reasoner (HCR) was evaluated using four different LLMs: Llama-4-Maverick-17B, GPT-4o-2024-11-20, Qwen2.5-VL-32B-Instruct, and Google Gemini 2.5 Flash. As described in Section 3.3, the HCR framework operated in a zero-shot manner on the test set. The LLMs were guided by our structured prompt to produce a detailed evaluation, from which a final quality classification was derived based on the synthesized score and predefined rules. Performance across all methods was assessed using accuracy (ACC), precision, recall, and F1-score (F1).

\begin{table*}[htbp]
	\centering
	\caption{Performance comparison of baseline vision models (trained) and our HCR with different LLMs (zero-shot) on the segmentation quality classification task. Best overall accuracy and precision are highlighted.}
	\label{tab:performance_results}
	\resizebox{0.8\textwidth}{!}{%
		\begin{tabular}{lcccc}
			\toprule
			\textbf{Model} & \textbf{Accuracy} & \textbf{Precision} & \textbf{Recall} & \textbf{F1-Score} \\
			\midrule
			\multicolumn{5}{l}{\textit{Baseline Vision Models (Trained on 383 samples)}} \\
			EfficientNet-B0 & 0.7188 & 0.6034 & \textbf{0.8974} & 0.7216 \\
			ResNet50 & 0.7292 & 0.6667 & 0.6667 & 0.6667 \\
			ViT-Base & 0.6979 & 0.7083 & 0.4359 & 0.5397 \\
			\midrule
			\multicolumn{5}{l}{\textit{HCR (LLMs - Zero-Shot Evaluation on 96 samples)}} \\
			Llama-4-Maverick & 0.6146 & \textbf{0.9565} & 0.3793 & 0.5432 \\
			GPT-4o & 0.6979 & 0.8718 & 0.5862 & 0.7010 \\
			Qwen2.5-VL-32B-Instruct & 0.6875 & 0.7188 & 0.7931 & 0.7541 \\
			Gemini-2.5-Flash & \textbf{0.7812} & 0.7937 & 0.8621 & \textbf{0.8264} \\
			\bottomrule
		\end{tabular}%
	}
\end{table*}

\subsection{Quantitative Performance}
The quantitative performance of the baseline vision models and our HCR approach utilizing different LLMs is summarized in Table~\ref{tab:performance_results}. The trained baseline models achieved varying levels of proficiency; for instance, ResNet50 attained an accuracy of 72.92\% and an F1-score of 0.6667, while EfficientNet-B0 recorded an accuracy of 71.88\% and a higher F1-score of 0.7216. In comparison, our HCR framework, operating without any task-specific training, demonstrated compelling performance. Notably, HCR when powered by Google Gemini 2.5 Flash achieved the highest accuracy (78.12\%) and precision (79.37\%) among all evaluated methods, with an F1-score of 0.8264. Other LLMs within the HCR framework also showed considerable capabilities; for example, GPT-4o yielded an accuracy of 69.79\% and an F1-score of 0.7010. These results indicate that LLMs, guided by the HCR's structured reasoning prompt, can provide quality assessments that are competitive with, and in certain aspects exceed, vision models specifically trained for this classification task.

\begin{figure*}[htbp]
	\centering
	\includegraphics[width=\textwidth]{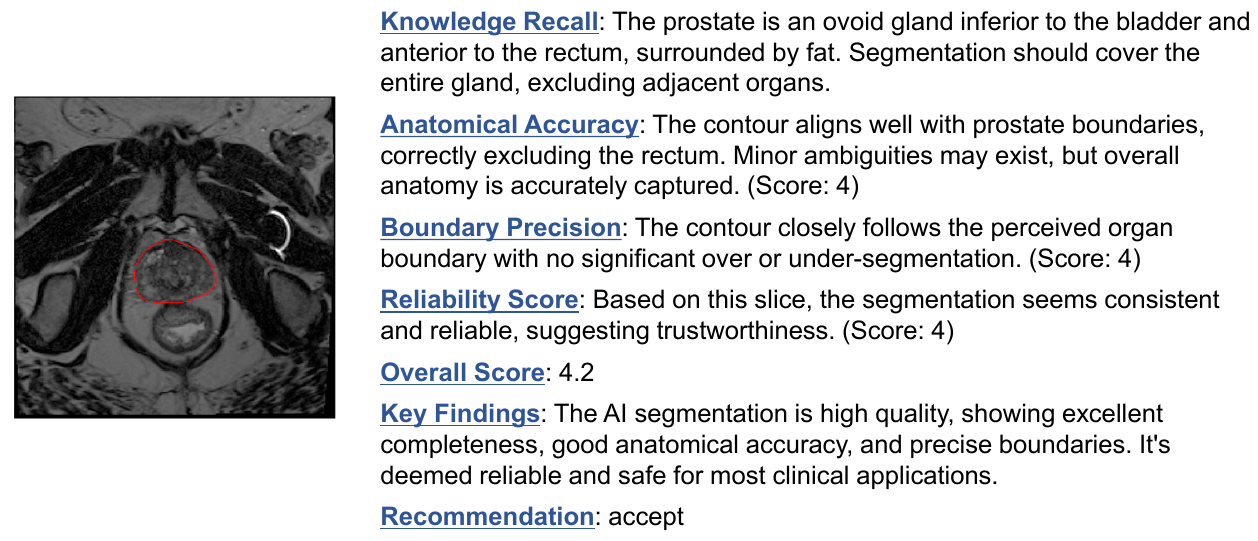}
	\caption{Illustrative example of the detailed clinical reasoning and structured output generated by different LLMs within the HCR framework for representative segmentation cases. These examples highlight the interpretability afforded by our approach.}
	\label{fig:llm_analysis_examples}
\end{figure*}

\subsection{Qualitative Analysis and Interpretability}
Beyond quantitative metrics, a key attribute of our HCR approach is its ability to generate detailed, interpretable textual explanations for its quality assessments. The structured output from HCR, detailing the LLM's reasoning across knowledge recall, visual feature analysis, anatomical inference, and clinical synthesis, provides valuable insights into the basis of its decision. Figure~\ref{fig:llm_analysis_examples} showcases examples of these comprehensive evaluations generated by different LLMs within the HCR framework, illustrating the depth of reasoning elicited by our prompting strategy. This interpretability offers a significant advantage over conventional classifiers and aligns with the increasing need for transparent AI in clinical settings.

\section{Discussion and Conclusion}
In this work, we introduced Hierarchical Clinical Reasoner (HCR), a framework leveraging LLMs to perform zero-shot, clinically-aligned quality assessment of medical image segmentations. Our results demonstrate that HCR, guided by a structured, multi-stage reasoning prompt, can achieve classification performance competitive with, and in some instances superior to, dedicated vision models trained for this task. This capability to provide detailed, interpretable evaluations moves beyond traditional pixel-agreement metrics and positions LLMs as effective clinical guardrails for enhancing the reliability of AI-driven segmentation in medical image analysis. A key aspect of HCR is its zero-shot application, obviating the need for extensive task-specific training data for the LLM evaluator itself. The structured reasoning pathway, from knowledge recall to clinical synthesis, allows HCR to produce assessments that are not only accurate but also provide a transparent rationale for its judgments. This articulated reasoning is a departure from many existing automated QC methods that primarily output quantitative scores or binary flags without detailed explanations.

Despite these promising results, our study has several limitations. The HCR framework's performance is intrinsically linked to the capabilities of the underlying LLM, and variations were observed across different models. While our structured prompting aims to ensure reliability, LLMs can occasionally produce inconsistent or unfaithful reasoning, necessitating careful model selection and potentially ensembling strategies for robust deployment. The generalizability of HCR to a broader range of imaging modalities, anatomical targets, and rarer segmentation error types not extensively covered in our current dataset requires further investigation. Moreover, like all AI systems, the potential for inherent biases within LLMs to affect evaluation fairness across different patient subgroups needs continued attention and mitigation strategies.

In conclusion, our Hierarchical Clinical Reasoner framework offers a promising new direction for the automated quality control of medical image segmentations. By guiding LLMs through a structured, multi-stage clinical reasoning process, HCR provides interpretable and clinically-aligned evaluations in a zero-shot manner, demonstrating performance comparable to trained specialist models. This work paves the way for leveraging the sophisticated reasoning abilities of LLMs to establish robust clinical guardrails, thereby fostering greater trust and reliability in the deployment of AI segmentation tools in healthcare.

\bibliography{mybibliography.bib}

\end{document}